\documentclass[12pt]{article}
\usepackage{amsmath, amssymb, amsthm}
\usepackage{setspace}
\usepackage{geometry}
\usepackage{graphicx}   
\usepackage{authblk}   
\usepackage{setspace}  
\usepackage{booktabs}
\usepackage{eurosym}

\title{Risk-Adjusted Policy Learning and the Social Cost of Uncertainty: Theory and Evidence from CAP evaluation}

\author[1]{Giovanni Cerulli}
\author[2]{Francesco Caracciolo}

\affil[1]{\footnotesize IRCrES-CNR, Research Institute on Sustainable Economic Growth, National Research Council, Italy}
\affil[2]{\footnotesize Department of Agricultural Sciences, University of Naples Federico II, Italy}

\date{\today}

\begin{document}
\maketitle

\begin{abstract}
\noindent
\footnotesize
This paper develops a risk-adjusted alternative to standard optimal policy learning (OPL) for observational data by importing Roy’s (1952) \emph{safety-first} principle into the treatment assignment problem. We formalize a welfare functional that maximizes the probability that outcomes exceed a socially required threshold and show that the associated pointwise optimal rule ranks treatments by the ratio of conditional means to conditional standard deviations. We implement the framework using microdata from the Italian \emph{Farm Accountancy Data Network} to evaluate the allocation of subsidies under the EU Common Agricultural Policy. Empirically, risk-adjusted optimal policies systematically dominate the realized allocation across specifications, while risk aversion lowers overall welfare relative to the risk-neutral benchmark, making transparent the \textit{social cost} of insurance against uncertainty. The results illustrate how safety-first OPL provides an implementable, interpretable tool for risk-sensitive policy design, quantifying the efficiency–insurance trade-off that policymakers face when outcomes are volatile.
\end{abstract}

\section{Introduction}

The problem of assigning individuals to treatments or policies on the basis of observed covariates has become central in econometrics, statistics, and machine learning (Dehejia, 2005; Hirano \& Porter, 2009; Cerulli, 2023). 
In the canonical framework of optimal policy learning (OPL), the planner is assumed to be risk-neutral and seeks to maximize expected social welfare under a treatment assignment rule (Manski, 2004; Kitagawa and Tetenov, 2018; Athey and Wager, 2021). 
Formally, this amounts to recommending the treatment with the highest conditional mean outcome for each individual (Bertsimas \& Kallus, 2020). 
While appealing in its simplicity, the risk-neutral approach disregards the variability of outcomes and hence abstracts from risk-sensitive concerns that are often critical in economic and social policy design.  

In contrast, the finance literature has long recognized the importance of risk-adjusted decision criteria. 
Roy’s (1952) seminal \emph{safety-first principle} advocates minimizing the probability that returns fall below a socially required threshold, thereby shifting attention from mean returns to risk-adjusted performance measures. 
Under normality, this principle yields a decision rule based on the ratio of expected returns to their standard deviations---an early precursor of the Sharpe ratio. 
The core insight is that decision-making under uncertainty should balance expected gains against the risk of undesirable outcomes (in particular, \textit{worst-case} scenarios).  

This paper transfers Roy’s safety-first principle into the domain of policy learning with observational data. 
We propose a \emph{risk-adjusted welfare functional}, defined as the probability that realized outcomes exceed a socially required minimum level, and show that it leads to a pointwise optimal policy based on the ratio of conditional means to conditional standard deviations of potential outcomes. 
Importantly, this mean–variance rule does not arise from quadratic utility or mean–variance preferences, but rather from a probabilistic safety criterion that places explicit weight on uncertainty.  

Our contributions are twofold. 
First, on the theoretical side, we establish the formal foundations of risk-adjusted OPL by deriving the safety-first welfare functional, its pointwise optimality, and its connections to location–scale families of distributions. 

Second, on the empirical side, we implement the framework using data from the \emph{European Farm Accountancy Data} (FADN),  to evaluate the allocation of subsidies under the European Union (EU) Common Agricultural Policy (CAP).  The CAP is the EU principal instrument for supporting agriculture, routinely absorbing close to one third of the EU budget (Espinosa et al., 2020). Within the CAP architecture, first‑pillar direct payments (DPs) are predominantly decoupled from production and are intended to stabilize farm incomes and, indirectly, safeguard food security (Biagini et al., 2020). Despite their fiscal centrality and broad coverage, the efficiency consequences of DPs remain contested. Because eligibility is largely defined by historical entitlements and land area—rather than performance-based criteria—the actual distribution of subsidies may diverge considerably from optimal allocations based on economic or social outcomes. In this context, OPL offers a promising framework for empirically evaluating whether the observed assignment of DPs could be improved through data-driven rules that account for heterogeneity in farm characteristics and performance.

\medskip
The remainder of the paper is organized as follows.
Section~\ref{sec:theory1} develops the OPL framework under risk neutrality.
Section~\ref{sec:theory2} introduces the \textit{safety-first} welfare objective derives the resulting mean–variance policy rule discussing its identification and interpretation.
Section~\ref{sec:app} presents the empirical setting, detailing the FADN microdata and the CAP institutional background.
Section~\ref{sec:res} presents the results of our analysis.
Section~\ref{sec:concl} concludes and discusses implications for risk-sensitive policy design.

\section{Literature}
\label{sec:lit} 

In environments characterized by uncertainty, the outcome of a given action depends not only on its expected return but also on the variability surrounding that return (Markowitz, 1952; Pratt, 1964). Thus, the decision to choose action $A$ over action $B$ cannot rely solely on comparing their average payoffs; it must also account for the degree of uncertainty each action entails. In other words, decision-making under uncertainty inherently involves trading off reward against risk.   

\begin{figure}[t]
\centering
\includegraphics[width=11cm]{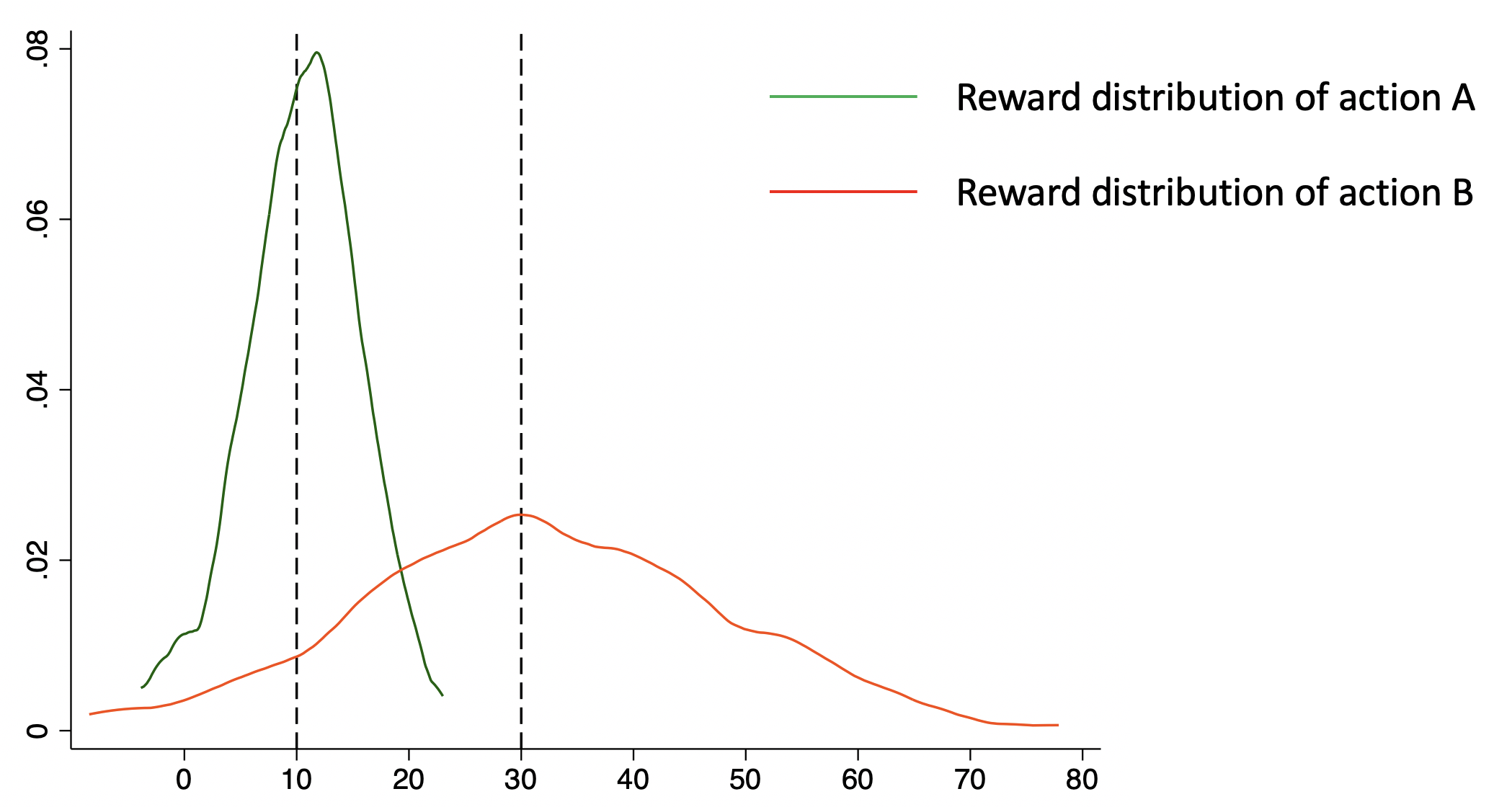}
\caption{\footnotesize Reward distributions for two actions, $A$ and $B$. Action $A$ delivers a lower mean return but with lower variability, while action $B$ yields a higher mean return accompanied by greater uncertainty.}
\label{fig:fig1}
\end{figure}

Figure \ref{fig:fig1} illustrates this trade-off. Action $A$ is safer, providing more stable outcomes at the cost of a lower mean return. Action $B$ promises higher returns but also exposes the decision-maker to larger fluctuations. In such cases, the optimal choice is not obvious: it depends on how returns and risks are jointly valued.  

This issue has been explicitly addressed in the multi-armed bandit literature that studies risk-sensitive agents. Instead of evaluating actions purely by their mean reward, these models incorporate the distributional properties of outcomes, such as their variance, when making choices (Sani et al., 2012). Once risk enters the objective function, classical algorithms designed to balance exploration and exploitation under risk neutrality may behave differently, and their asymptotic performance can diverge from the standard case.  

In the context of optimal policy learning (OPL) with observational data (see Sani et al., 2012; Chandak et al., 2021; Cassel et al., 2023), a growing body of work has focused on estimating the variance of policy returns. For example, Chandak et al. (2021) propose a consistent estimator of the variance of the return under a policy $\pi$:  
\begin{equation} \label{eq:var1}
\sigma^{2}(\pi) = \text{Var}[Y(\pi(X))].
\end{equation}
This formulation emphasizes that outcome distributions are defined not only by their mean but also by the variability around it.  

A recent contribution by Osama et al.\ (2020) develops a robust policy learning framework 
that explicitly targets tail risks in safety-critical applications. Instead of minimizing the 
expected cost, their method aims to minimize the $(1-\alpha)$-quantile of the cost distribution: 
\[
\pi^\ast = \arg \min_{\pi \in \Pi} \inf \{ y_\alpha : P_\pi(y \leq y_\alpha) \geq 1-\alpha \},
\]
thus prioritizing robustness against extreme outcomes. Building on conformal prediction, 
the approach provides finite-sample statistical guarantees and is particularly suited to 
settings with limited overlap across features and treatments. Empirically, their robust 
policy reduces the probability of adverse tail events relative to both past and mean-optimal 
policies.

This study proposes a complementary yet conceptually different risk-adjusted framework. Importing Roy's (1952) safety-first principle into the OPL setting, 
we define social welfare as the probability that outcomes exceed a socially required 
threshold and derive a pointwise optimal policy based on the ratio of conditional means to 
conditional standard deviations:
\[
\pi(x) = \arg \max_{t \in T} \frac{\mu_t(x)}{\sigma_t(x)}.
\]
This ``mean-to-risk'' rule, reminiscent of the Sharpe ratio, highlights the 
efficiency-insurance trade-off faced by policymakers: risk-averse policies reduce exposure 
to downside variability at the cost of lower expected welfare relative to the risk-neutral 
benchmark. 

Heuristically, this decision rule was proposed by Cerulli (2024). Rather than focusing on the \textit{overall} policy variance (Eq.~\ref{eq:var1}), he advocates for analyzing the \textit{conditional} variance, thereby allowing explicit modeling of agents’ risk preferences. Conditional uncertainty can be measured as:  
\begin{equation} \label{eq:variance}
\text{Var}(Y|X) = E[Y^2|X] - (E[Y|X])^2.
\end{equation}
Accordingly, for observation $i$ and action $a$, the conditional variance is:
\begin{equation} \label{eq:variance2}
\hat{\sigma}^{2}_{i}(a,X_{i}) = \hat{E}(Y^{2}_{i}|A_{i}=a,X_{i}) - \hat{E}(Y_{i}|A_{i}=a,X_{i})^2,
\end{equation}
where the conditional expectations on the right-hand side can be estimated via machine learning methods. Hence, the characterization of each action relies on the pair:
$$\big[ \hat{\mu}_{i}(a,X_{i}),\; \hat{\sigma}_{i}(a,X_{i}) \big],$$
with $\hat{\sigma}_{i}(\cdot)$ denoting the estimated conditional standard deviation.  

In a risk-neutral framework, decision-makers ignore $\hat{\sigma}_{i}$ and simply select the action with the highest conditional mean. By contrast, under risk aversion, agents favor actions that provide a more favorable trade-off between mean return and uncertainty. This preference can be represented through utility functions that depend jointly on conditional means and standard deviations.Two specifications are:  

\noindent
\textit{Linear risk-averse preferences}:  
\begin{equation} \label{eq:linear_pref1}
U_{i,L} = \frac{\hat{\mu}_{i}}{\hat{\sigma}_{i}},
\end{equation}
which yields linear indifference curves of the form $\hat{\mu}_{i} = k \cdot \hat{\sigma}_{i}$.  

\noindent
\textit{Quadratic risk-averse preferences}:  
\begin{equation} \label{eq:quadratic_pref1}
U_{i,Q} = \frac{\hat{\mu}_{i}}{\hat{\sigma}_{i}^{2}},
\end{equation}
which implies quadratic indifference curves $\hat{\mu}_{i} = k \cdot \hat{\sigma}_{i}^{2}$.  

\noindent
Importantly, different specifications may rank actions differently. For instance, action $A$ may dominate $B$ under linear preferences but not under quadratic ones, highlighting how the assumed structure of risk aversion critically shapes policy recommendations (Cerulli, 2024). The key implication is that risk-sensitive OPL may select actions that deviate from the first-best, risk-neutral allocation. This deviation can be measured through a new policy functional to maximize, such as:  
\begin{equation} \label{eq:LRApolicy1}
\pi^{\text{LRA}}(X) = \arg\max_{a \in \mathcal{A}} \frac{\mu(a,X)}{\sigma(a,X)},
\end{equation}
with:  
\begin{equation} \label{eq:LRApolicy2}
V(\pi^{\text{LRA}}(X)) \leq V(\pi^{\text{FB}}(X)),
\end{equation}
and welfare loss (or regret) defined as:  
\begin{equation} \label{eq:LRAregret}
R = V(\pi^{\text{FB}}(X)) - V(\pi^{\text{LRA}}(X)).
\end{equation}
This framework makes explicit how the introduction of risk preferences can alter the ranking of actions, the resulting policy rules, and the corresponding welfare outcomes. It also clarifies that prudence in the face of uncertainty—while rational from the standpoint of a risk-averse agent—comes at the cost of welfare loss relative to the risk-neutral benchmark.
The next two sections provide a formal derivation of (\ref{eq:LRApolicy1}), rooted in Roy's \textit{safety-first} decision principle.

\section{First-best policy in a risk-neutral setting}
\label{sec:theory1}

Let $\mathcal T=\{0,1,\dots,M-1\}$ denote a finite set of mutually exclusive treatments, 
$X\in\mathcal X$ a vector of pre-treatment covariates, and 
$\{Y_t : t\in\mathcal T\}$ the potential outcomes associated with each treatment 
(Holland, 1986). A policy is a measurable function: 
\[
\pi:\mathcal X \to \mathcal T,
\]
mapping individuals' covariates to a treatment recommendation. 
Following the canonical definition in the policy learning literature 
(Zhou, Athey, \& Wager, 2023; Athey and Wager, 2021; Kitagawa and Tetenov, 2018), 
the welfare generated by a policy $\pi$ under risk neutrality is:
\begin{equation}
W(\pi) \;=\; \mathbb E\!\left[\,Y_{\pi(X)}\,\right].
\end{equation}
\noindent
By the law of iterated expectations, we have that:
\begin{equation}
W(\pi) 
= \mathbb E\!\Big[\, \mathbb E[Y_{\pi(X)} \mid X] \,\Big] 
= \mathbb E\!\Big[\, \mu_{\pi(X)}(X) \,\Big],
\end{equation}
where $\mu_{\pi(X)}(X)$ is the conditional mean potential outcome corresponding to $t=\pi(X)$.
\noindent
Maximizing $W(\pi)$ over all measurable policies amounts to solving a 
pointwise optimization problem in $x$. 
Indeed, for any fixed $x\in\mathcal X$ and any $t\in\mathcal T$,
\[
\mu_{\pi(x)}(x) \;\le\; \max_{t\in\mathcal T}\mu_t(x),
\]
with equality if and only if $\pi(x)\in\arg\max_{t\in\mathcal T}\mu_t(x)$. 
Therefore, the \emph{first-best} or \emph{oracle policy} under risk neutrality is
\begin{equation}
\pi^\star(x)\;\in\;\arg\max_{t\in\mathcal T}\,\mu_t(x).
\end{equation}

The corresponding maximal welfare level is
\begin{equation}
W^\star \;=\; \mathbb E\!\left[\max_{t\in\mathcal T}\mu_t(X)\right].
\end{equation}

The policy $\pi^\star$ is referred to as \emph{first-best} and to be feasible assumes that the planner has perfect knowledge of the conditional mean potential outcomes $\{\mu_t(x): t\in\mathcal T\}$ for every possible $x$. 
In this idealized benchmark, the planner can assign each individual 
to the treatment that maximizes their expected outcome, thus achieving the highest possible social welfare. 

Estimating the first-best policy, however, requires counterfactual knowledge since, for each individual, only one potential outcome $Y_t$ is observed, while all others remain counterfactual (Holland, 1986; Rubin, 1974). 
This missing-data structure prevents direct observation of $\mu_t(x)$. 
To overcome this, identification requires assumptions. 
A standard framework is based on \emph{unconfoundedness} and \emph{overlap} 
(Rosenbaum and Rubin, 1983): 
\[
\{Y_t : t\in\mathcal T\} \;\perp\!\!\!\perp\; T \mid X
\quad\text{and}\quad 
0<\Pr(T=t\mid X=x)<1 \;\;\forall t\in\mathcal T, \;x\in\mathcal X,
\]
where $T$ denotes the realized treatment. 
Under these conditions, the conditional mean $\mu_t(x)=\mathbb E[Y_t\mid X=x]$ 
is identified from observable data as $\mathbb E[Y\mid T=t, X=x]$. 
This makes it possible to estimate $\mu_t(x)$ using regression, 
machine learning, or semiparametric methods, 
and then construct a plug-in estimate of $\pi^\star$. 

Thus, the first-best serves as a normative benchmark---what the planner would do if the conditional potential outcome means were known. 
Practical policy learning methods aim to approximate this ideal as closely as possible from finite samples of observed data.

\section{The \textit{safety-first} welfare optimization criterion} \label{sec:theory2}

As seen above, in standard optimal policy learning (OPL), the planner is assumed to be risk-neutral. By assuming a risk-averse planner, in this section we derive a risk-adjusted alternative optimal policy based on the ratio of conditional means to conditional standard deviations of potential outcomes, showing that this rule arises from a safety-first optimization.

Given a threshold $y_\star\in\mathbb R$ (typically $y_\star>0$), we define the \emph{safety-first welfare} as:
\[
W(\pi)\;:=\;\mathbb{E}\big[\mathbf{1}\{Y_{\pi(X)}\ge y_\star\}\big],
\]
where the policy $\pi$ is a measurable function $\pi:\mathcal X\to\mathcal T$.  
For $x\in\mathcal X$ and $t\in\mathcal T$ let
\[
h(x,t)\;:=\;\Pr\!\big(Y_t\ge y_\star\mid X=x\big).
\]
By the law of iterated expectations,
\[
W(\pi)\;=\;\mathbb{E}\!\left[h\big(X,\pi(X)\big)\right].
\]

\paragraph{Lemma} (\textit{Pointwise optimality}).
Under standard measurability assumptions, it holds that
$$
\sup_{\pi}\;\mathbb{E}\!\left[h\big(X,\pi(X)\big)\right]
\;=\;
\mathbb{E}\!\left[\max_{t\in\mathcal T} h\big(X,t\big)\right]$$
and the optimal policy is:
$$\pi^\star(x)\in\arg\max_{t\in\mathcal T} h(x,t)\; \text{ a.s.}$$
\noindent
\emph{Proof.} \\ 
(i) \emph{(inequality $\le$)} For any policy $\pi$ and almost every $X$, one has $h\!\big(X,\pi(X)\big)\le \sup_{t\in\mathcal T} h(X,t)$. Taking expectations yields:
\[
\sup_{\pi}\mathbb{E}\!\left[h\big(X,\pi(X)\big)\right]
\;\le\;
\mathbb{E}\!\left[\sup_{t\in\mathcal T} h(X,t)\right].
\]
(ii) \emph{(attainability)} Since $\mathcal T$ is finite, the set
$A_t:=\{x: h(x,t)=\max_{s\in\mathcal T}h(x,s)\}$ is measurable for each $t$.  
Fix a deterministic tie-breaking rule (e.g., the smallest index) and define $\pi^\star(x)$ as any measurable selection in $\arg\max_{t\in\mathcal T}h(x,t)$. Then $h\!\big(X,\pi^\star(X)\big)=\max_{t\in\mathcal T}h(X,t)$ a.s., so that
\[
\mathbb{E}\!\left[h\big(X,\pi^\star(X)\big)\right]
=
\mathbb{E}\!\left[\max_{t\in\mathcal T} h(X,t)\right].
\]
Combining (i) and (ii) proves the claim. \hfill$\square$

\medskip

\paragraph{Corollary} (\textit{Normality and the $z$-score}).
If for each $t$ it holds that $Y_t\mid X\sim\mathcal N\!\big(\mu_t(X),\sigma_t^2(X)\big)$, then:
\begin{eqnarray}
h(x,t) 
&=& \Pr\!\left(Y_t \ge y_\star \,\big|\, X=x \right) \nonumber \\[6pt]
&=& \Pr\!\left( \frac{Y_t - \mu_t(x)}{\sigma_t(x)} 
      \;\ge\; \frac{y_\star - \mu_t(x)}{\sigma_t(x)} \,\Big|\, X=x \right) \nonumber \\[6pt]
&=& 1 - \Phi\!\left( \frac{y_\star - \mu_t(x)}{\sigma_t(x)} \right) \nonumber \\[6pt]
&=& \Phi\!\left( \frac{\mu_t(x) - y_\star}{\sigma_t(x)} \right).
\end{eqnarray}

\paragraph{The $\mu/\sigma$ rule.}
If $y_\star=0$ (or equivalently if outcomes are centered: $Y'_t:=Y_t-y_\star$), the optimal policy coincides with
\[
\pi^\star(x)\in\arg\max_{t\in\mathcal T}\;
\frac{\mu_t(x)}{\sigma_t(x)}.
\]
In other words, the safety-first policy selects, pointwise in $x$, the treatment with the highest \emph{signal-to-noise ratio} (the $z$-score relative to the threshold). The Generalization to common location–scale families is straightforward. Suppose that:
$$Y_t\mid X \stackrel{d}{=} \mu_t(X)+\sigma_t(X)\,\varepsilon,$$
with $\varepsilon$ independent of $t$, and cdf $F(\cdot)$ fixed and strictly increasing, then we have that:
\[
h(x,t)=\Pr\!\big(\mu_t(x)+\sigma_t(x)\varepsilon\ge y_\star\mid X=x\big)
=\Pr\!\left(\varepsilon\ge \frac{y_\star-\mu_t(x)}{\sigma_t(x)}\right)
=1-F\!\left(\frac{y_\star-\mu_t(x)}{\sigma_t(x)}\right).
\]
Since $F(\cdot)$ is increasing, for $t_1,t_2\in\mathcal T$ we have:
\[
h(x,t_1)\ge h(x,t_2) 
\;\Longleftrightarrow\;
\frac{y_\star-\mu_{t_1}(x)}{\sigma_{t_1}(x)}
\;\le\;
\frac{y_\star-\mu_{t_2}(x)}{\sigma_{t_2}(x)}
\;\Longleftrightarrow\;
\frac{\mu_{t_1}(x)-y_\star}{\sigma_{t_1}(x)}
\;\ge\;
\frac{\mu_{t_2}(x)-y_\star}{\sigma_{t_2}(x)}.
\]
Therefore, the maximizer remains $\arg\max_t (\mu_t-y_\star)/\sigma_t$.

\subsection{Simulation}

To better position and motivate the contribution of this paper, it is useful to provide a simple simulation that makes explicit how the social cost of uncertainty arises. To this end, we compare three (optimal) policy rules:

\begin{itemize}
    \item oracle policy:
    \begin{equation}
        \pi^{OR}(X) = \mathbf{1}\!\big[(Y_{1}\mid X) - (Y_{0}\mid X) > 0 \big],
        \label{eq:oracle}
    \end{equation}

    \item risk-neutral first-best policy:
    \begin{equation}
        \pi^{RN}(X) = \mathbf{1}\!\big[\mu_{1}(X) - \mu_{0}(X) > 0 \big],
        \label{eq:risk_neutral}
    \end{equation}

    \item risk-averse linear first-best policy:
    \begin{equation}
       \pi^{RA}(X)=\mathbf{1}\!\left[ \frac{\mu_{1}(X)}{s_{1}(X)} > \frac{\mu_{0}(X)}{s_{0}(X)} \right],
        \label{eq:risk_averse}
    \end{equation}
\end{itemize}

The last two expressions are the focus of our previous comparison. The first expression defines the \emph{oracle policy}, which serves as the theoretical benchmark: it assigns treatment $1$ rather than $0$ whenever the individual causal effect $Y_{1}-Y_{0}$ is strictly positive (possibly conditional on $X$). This rule is \emph{omniscient} because it requires knowledge of the realizations of the \emph{individual potential outcomes}, which are never jointly observable in practice. Although the oracle policy is not implementable, in a simulation setting - where both $Y_{1}$ and $Y_{0}$ can be made simultaneously known - it provides the conceptual ideal against which feasible (i.e., estimable) policies can be evaluated.

These three rules thus highlight the tension between theoretical optimality and practical implementability. The oracle policy defines the true but unattainable optimum. The risk-neutral first-best is the standard implementable approximation (under unconfoundedness and overlapping), yet it may diverge substantially from the oracle in the presence of high uncertainty. In such contexts, a risk-averse policy may be socially preferable: although “suboptimal” in terms of expected welfare, it curbs erratic realizations and protects against \emph{worst-case} scenarios. Indeed, if the worst-case outcome under the risk-neutral first-best materializes, realized welfare may fall well below that delivered by a more conservative but stable assignment rule. We operationalize this argument below using a simple data-generating process.

\smallskip 
\smallskip 

\begin{figure}[htbp]
    \centering
    \includegraphics[width=1\textwidth]{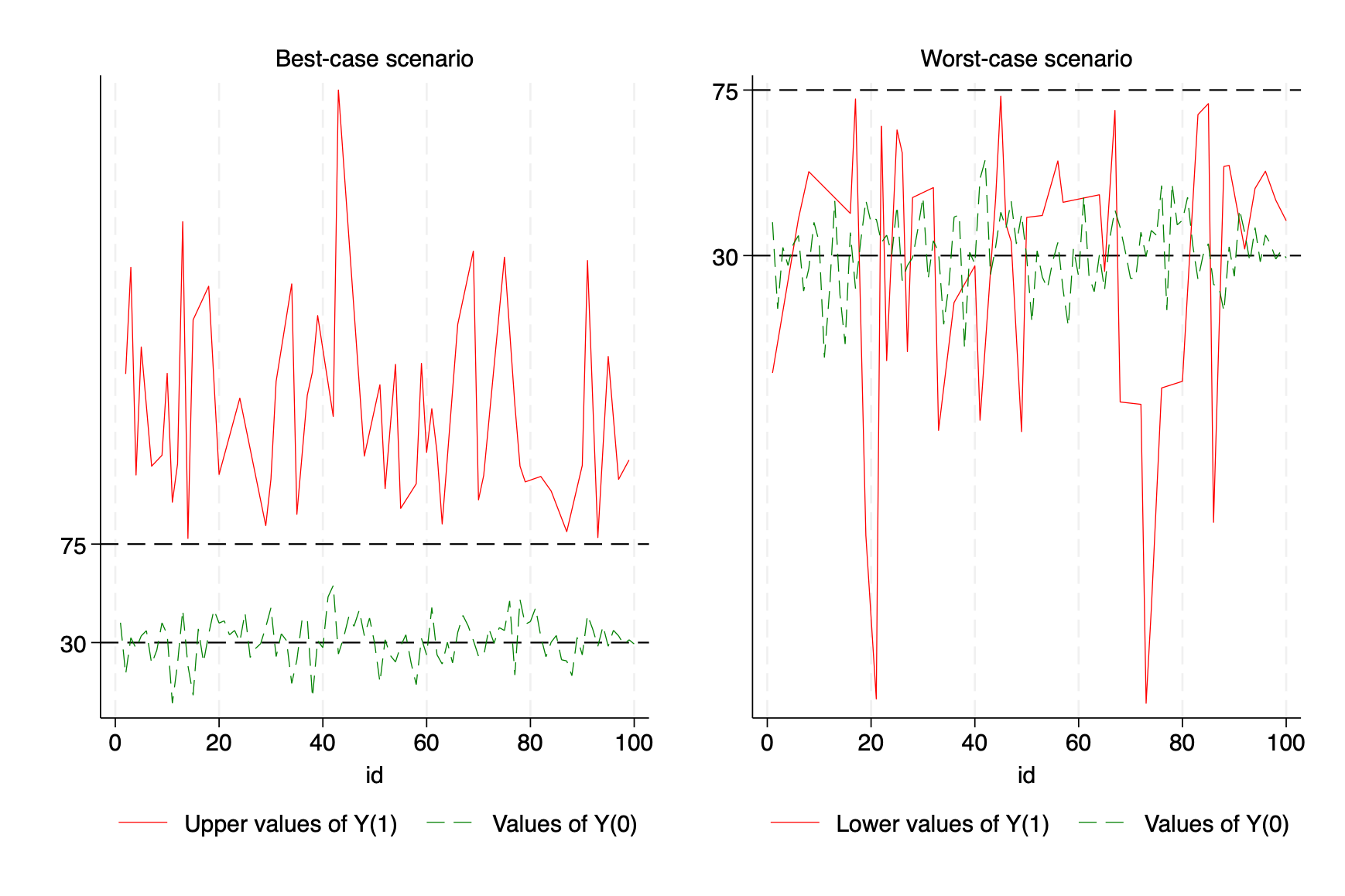}
    \caption{\footnotesize Text}
    \label{fig:gr_sim}
\end{figure}

\noindent
\textit{Data Generating Process}. We consider a binary treatment setting, with $T \in \{0,1\}$, and corresponding potential outcomes:
\[
Y_{0i} \sim \mathcal{N}(\mu_0,\sigma_0^2),\qquad
Y_{1i} \sim \mathcal{N}(\mu_1,\sigma_1^2).
\]
\noindent
We fix the parameters of the two distributions as:
\[
\mu_0 = 30,\qquad \mu_1 = 75,\qquad \sigma_0 = 10,\qquad \sigma_1 = 65.
\]
and draw 100 independent values from these distributions. Then, we compute the risk-neutral decision rule. As $\mu_1=75>\mu_0=30$, according to this rule we should always treat in each run. Conversely, according to the risk-adjusted policy rule, $\mu_1/\sigma_1\approx 1.154 < \mu_0/\sigma_0 = 3$, leading us to choose not to treat in all runs. When it comes to the oracle decision rule, however, we might have a mixed situation, where sometimes it is optimal to treat and sometimes not, with $\pi^{RN}\neq\pi^{RA}\neq\pi^{OR}$, depending on the erratic realizations of both $Y_1$ and $Y_0$. This means that — if we knew the actual realizations $Y_1$ and $Y_0$ - it would be better \textit{sometimes} to treat and \textit{sometimes} not to treat. The realized outcomes under each policy $P\in\{RN, RA, OR\}$ are:
\[
Y^{(P)}_i \;=\; Y_{0i} + \pi^{P}_i\,(Y_{1i}-Y_{0i})
\;=\; \pi^{P}_i\,Y_{1i} + \bigl(1-\pi^{P}_i\bigr)Y_{0i}.
\]
and the corresponding welfare can be simply estimated by the sample mean as:
\[
W(P) \;=\; \frac{1}{n}\sum_{i=1}^n Y^{(P)}_i,
\qquad P\in\{RN,RA,OR\}.
\]
where $W(RN)=78$ and $W(RA)=30$. 
Figure~\ref{fig:gr_sim} sets out the results of our DGP simulation. We compare the pattern of $Y_0$ — the outcome under the risk-adjusted policy — with the \emph{best} and \emph{worst} realizations of $Y_1$. In our simulation, the best scenario occurs whenever $Y_1>75$, while the worst scenario occurs when $Y_1<75$. In the best case, the erratic nature of $Y_1$ sometimes yields values well above its mean ($75$), making treatment uniformly optimal. By contrast, in the worst case — where volatility drives $Y_1$ far below $75$ — the picture is mixed, with many runs in which it is preferable not to treat. If this worst-case scenario materializes, the loss from assigning treatment can even exceed the gap from forgoing the always-treat rule (i.e., $75-30=45$), since $Y_1$ may fall substantially below $Y_0=30$. 

This prospect is particularly concerning for a risk-sensitive decision maker and naturally motivates a risk-adjusted strategy. In short, avoiding costly worst-case outcomes is precisely why risk aversion arises in policy decisions contexts.

\section{Application to the Common Agricultural Policy}
\label{sec:app} 

In this section, we provide an empirical application relying on an unbalanced longitudinal dataset derived from the \textit{Farm Accountancy Data Network}, comprising 9,336 Italian arable farms and a total of 31,866 observations from 2010 to 2022 (European Commission, 2023). From a policy assessment perspective, FADN data is particularly valuable due to its detailed and statistically representative coverage of farm structures, production systems, costs, revenues, and subsidies. The dataset includes detailed variables on land use (e.g., Utilized Agricultural Area – UAA, rented vs. owned land), crop and livestock activities (e.g., crop areas, animal numbers, production volumes), labor (both family and hired labor, measured in annual work units), and machinery use. Moreover, the dataset provides comprehensive information on production costs (e.g., inputs, maintenance, depreciation), revenues (by product category), CAP subsidies (both decoupled and coupled payments), investments (e.g., in buildings, equipment, land), and overall farm income. This enables policymakers and researchers to evaluate the impacts of agricultural and rural policies—such as CAP—at the farm level, allowing for ex-ante and ex-post assessments.

The \textit{CAP} consists of two main components, or "pillars": the First Pillar, which includes direct payments and market measures, and the Second Pillar, which supports rural development through co-financed programs. While both pillars are central to the EU’s agricultural strategy, this paper focuses exclusively on First Pillar subsidies, and in particular, on decoupled direct payments.

There are three main reasons for this choice. First, direct payments under the First Pillar are allocated according to highly mechanical rules, primarily based on historical entitlements and eligible hectares of UAA, with only minimal conditionality attached. As such, they provide an ideal test case for evaluating the upper bound of potential efficiency gains from adopting data-driven selection rules. In contrast to the Second Pillar — where funds are already distributed through competitive procedures and eligibility criteria — the First Pillar remains a system, largely untargeted and undifferentiated. If OPL can demonstrate substantial welfare improvements in this context, where no selection mechanism is currently in place, this would underscore its practical value for policy innovation.

Second, the First Pillar accounts for the  majority of CAP expenditures, routinely absorbing more than 50\% of the CAP budget in Italy (Pierangeli et al., 2025).  First Pillar subsidies are  highly concentrated and controversial from an equity perspective. Historically 20\% of beneficiaries received 80\% of payments and this pattern has been remarkably persistent across countries and CAP reform cycles (Dinis, 2024).

Third, limiting the scope to First Pillar support allows us to work with a relatively homogeneous treatment assignment structure, improving interpretability and internal validity. While rural development measures under the Second Pillar are of great policy relevance, they involve highly heterogeneous instruments and objectives, often tied to specific regional, environmental, or structural considerations. These features, while important, make the Second Pillar less suitable for illustrating the general value of OPL as a policy evaluation and design tool.

Nearly all farms in our dataset are beneficiaries of Pillar~I subsidies, while about one third also receive Pillar~II payments. The treatment variable is defined as the amount of standardized subsidies per hectare of UAA, divided into three categories (category 0, category 1, and category 2) representing quantiles corresponding to \emph{low} (0), \emph{medium} (1), and \emph{high} (2) levels of support.  

For the purpose of this application, we focus on arable crops farms, both individually and in aggregate. These farm types are highly representative of Italian extensive agriculture and are particularly relevant for the analysis of CAP support.  

Several variables have been used in the literature to measure farming outcomes, including \textit{net farm output}, \textit{net income}, \textit{operating income,} and \textit{gross saleable production} (Carillo et al., 2017). These indicators capture related aspects of farm performance. \textit{Net farm output} reflects the gross value of agricultural production (excluding subsidies), net of specific variable costs, and thus represents the economic yield of farming activities. Operating income includes broader operational costs, while net income also accounts for subsidies, taxes, and financial charges. Empirical evidence shows strong correlations among these measures, with pairwise coefficients exceeding 93\%, indicating they convey similar information on profitability and economic performance. Given their interchangeability, this study adopts \textit{net farm output} as the primary outcome variable.

In addition to the treatment and outcome variables, the analysis controls for the principal factors of production. 
These include \textit{machinery capital} (expressed in kilowatts), \textit{labor units}, \textit{fixed costs}, and \textit{variable costs}. Together, these variables provide a comprehensive representation of the main inputs employed in agricultural production. By accounting for them, we capture heterogeneity in resource endowments and production intensity across farms, thus reducing potential confounding and improving the reliability of causal inference.  
 
Finally, since \emph{``arable crops''} are extensive farming systems, all monetary variables -- subsidies, outcomes, and input factors -- are standardized per hectare of UAA. 
This adjustment ensures comparability across farms of different sizes 
and allows us to interpret treatment and outcome measures in intensity terms rather than absolute levels.

\section{Results}
\label{sec:res}

Table~\ref{tab:descriptive} reports summary statistics for the  variables used in the analysis. The average net farm output amounts to euro 1,540 per hectare, though with considerable variation across farms, (SD: euro 3,490, Max: $\approx$ euro 55,300). Machine power and labor intensity also show substantial heterogeneity, with averages of 11.68 kW/ha and 0.095 LU/ha, respectively. Both fixed and variable costs exhibit high dispersion, reflecting structural diversity within the sample. Regarding policy variables, 35.5\% of farms received payments under the Second Pillar, while only 2.7\% benefited from national subsidies.

Tables~\ref{tab:main_results_tabN}--\ref{tab:main_results_tabQ} report the core empirical findings obtained under the three different risk preference specifications considered in this study: risk-neutral, risk-averse linear, and risk-averse quadratic \footnote{All computations in this paper were carried out using the Stata command \texttt{opl\_ma\_fb}, developed by Cerulli (2025b). See also Cerulli (2025a).}. Each table is structured in a consistent manner, detailing the composition of the dataset, the policy information, the empirical distribution of the actions, and the estimated value-functions under both actual and (counterfactual) optimal policies.

Across all three specifications, the dataset comprises 31,866 training observations, with no additional new data provided. We are not interested in fact in projecting the optimal welfare over ``new'' units.  This implies that all welfare estimates and counterfactual exercises are related to the training data. 

The treatment assignment problem is defined over three discrete actions $\{0,1,2\}$, referring to low ($0$), medium ($1$), and high ($2$) CAP's first pillar provisions to farms.  

The optimal policy is estimated using the same set of covariates (\textit{machinery capital}, \textit{labor units}, \textit{fixed and variable costs}, and \textit{other subsidy} indicators, including CAP's second pillar, and national support). 

Since we use three equidistant quantiles, the action frequencies are highly balanced, with each treatment capturing roughly one third of the sample (34.1\% for action 0, 32.9\% for action 1, and 32.9\% for action 2), ensuring balanced distribution mitigating concerns about lack of support. We provide now a more detailed comment of each table. 

\vspace{0.5cm}
\noindent
\textit{Risk-neutral specification}.
In the risk-neutral case (Table~\ref{tab:main_results_tabN}), the estimated value-function of the actual policy reaches 11.68, whereas the counterfactual optimal policy attains a substantially higher value of 13.63. The relative difference underscores the inefficiency of the observed assignment and the welfare gains that could be realized through policy re-optimization. Importantly, the rate of matches between the actual and the optimal policy is only 0.31, confirming that the observed treatment assignments are often inconsistent with what would be recommended under the risk-neutral welfare criterion. This highlights that in practice, treatment decisions deviate markedly from what would maximize expected welfare.

\vspace{0.5cm}
\noindent
\textit{Risk-averse linear specification}.
When risk aversion is incorporated through a linear specification (Table~\ref{tab:main_results_tabL}), the overall magnitude of welfare declines, reflecting the penalty associated with variability in outcomes. The actual policy delivers a value-function of 7.80, while the optimal counterfactual improves this outcome to 11.16. Although the level of welfare is reduced compared to the risk-neutral case, the relative improvement of the optimal over the actual policy remains large. The rate of optimal policy matches again equals 0.31, indicating that the inefficiency of the observed allocation persists independently of the assumed preference structure. This suggests that suboptimality is structural rather than preference-driven.

\vspace{0.5cm}
\noindent
\textit{Risk-averse quadratic specification}.
Under the quadratic risk-averse framework (Table~\ref{tab:main_results_tabQ}), the emphasis on downside protection becomes even more evident. The realized policy achieves a value of only 6.34, whereas the optimal policy increases this to 11.12. Although the optimal value-function is slightly lower than in the linear risk-averse case, it is still markedly higher than the realized allocation, confirming the robustness of welfare gains from re-optimization. The persistence of a low match rate (0.31) once again reveals that actual treatment assignments diverge systematically from counterfactual optima, even when the evaluation criterion is highly conservative.

\vspace{0.5cm}
\noindent
Three general insights emerge from the comparison across specifications. First, in all cases the optimal policy dominates the realized allocation, underscoring the potential for welfare improvements through data-driven policy design. Second, the absolute magnitude of welfare gains depends on the degree of risk aversion: the risk-neutral setting delivers -- as must be -- the highest expected welfare, while quadratic risk aversion delivers the lowest, in line with theory. This highlights the importance of carefully aligning the policy evaluation criterion with the policymaker’s tolerance for risk. Third, the consistently low rate of agreement between actual and optimal assignments suggests that inefficiency in the real-world allocation process is pervasive and not sensitive to the assumed preference structure.

Overall, the evidence from these Tables points to the existence of systematic \textit{welfare losses} under current allocation rules, but also emphasizes the role of data-driven optimal policies in delivering improved outcomes under alternative risk preferences. This constitutes strong support for the use of policy learning methods with observational data as a tool to inform more efficient and context-sensitive decision-making. In particular, the reduced level of welfare induced by risk-averse preferences reflects the \textit{social cost} associated with policymakers’ need to safeguard against uncertainty. 
Such cost arises because risk-averse criteria deliberately sacrifice some expected welfare in exchange for a lower exposure to variability, thereby prioritizing stability over maximal returns. 
This conservative stance can be interpreted as a rational response to the fear of\textit{ worst-case} scenarios, yet it highlights the trade-off faced by policymakers between efficiency and protection against downside risks.

\noindent
\begin{table}[htbp]
\centering
\caption{Descriptive Statistics of Main Variables}
\begin{tabular}{lrrrr}
\toprule
\textbf{Variable} & \textbf{Mean} & \textbf{SD} & \textbf{Min} & \textbf{Max} \\
\midrule
\textit{Continuous variables} \\
Net farm output (1,000 euro/ha)      & 1.54    & 3.49    & 0.00004 & 55.33 \\
Machine power (kW/ha)             & 11.68   & 18.30   & 0.17    & 444.55 \\
Labour units (LU/ha)              & 0.095   & 0.165   & 0.0008  & 2.29 \\
Fixed costs (1,000 euro/ha)                & 0.54  & 1.12 & 0       & 28.04 \\
Variable costs (1,000 euro/ha)             & 1.55  & 2.69 & 0       & 73.59 \\
\midrule
\textit{Dummy variables (mean = share)} \\
II Pillar beneficiary = 1         & 0.355   & 0.479   & 0       & 1 \\
National subsidies beneficiary = 1 & 0.027  & 0.163   & 0       & 1 \\
\midrule
\textit{First Pillar terciles (euro/ha)} & \multicolumn{4}{c}{} \\
Tercile 1 (Low)                   & 34.13\% &         &         &  \\
Tercile 2 (Mid)                   & 32.93\% &         &         &  \\
Tercile 3 (High)                  & 32.94\% &         &         &  \\
\bottomrule
\end{tabular}
\label{tab:descriptive}
\end{table}

\begin{table}[htbp]
\centering
\caption{Main Results: Risk-neutral}
\label{tab:main_results_tabN}
\scriptsize
\begin{tabular}{lcl}
\hline
\multicolumn{3}{c}{\textbf{Data Information}} \\
\hline
Number of training observations                    & = & 31,866  \\
Number of used training observations (optimal policy)     & = & 31,866  \\
Number of used training observations (non-optimal policy) & = & .   \\
Number of new observations                         & = & 0  \\
Number of used new observations (optimal policy)   & = & 0  \\
Number of used new observations (non-optimal policy) & = & .   \\
\hline
\multicolumn{3}{c}{\textbf{Policy Information}} \\
\hline
Target variable        & : & net farm output \\
Features               & : & Labor, fixed costs, variable costs, \\
                       &   & subsidy (pillar 2), subsidy (national) \\
Policy variable        & : & treat \\
Number of actions      & = & 3 \\
Actions                & = & \{0, 1, 2\} \\
\hline
\multicolumn{3}{c}{\textbf{Frequencies of the actions (training data)}} \\
\hline
Action 0 & Freq. = 10,876 & Percent = 34.13 \\
Action 1 & Freq. = 10,494 & Percent = 32.93 \\
Action 2 & Freq. = 10,496 & Percent = 32.94 \\
Total    & Freq. = 31,866 & Percent = 100.00 \\
\hline
\multicolumn{3}{c}{\textbf{Training Data Results}} \\
\hline
Value-function of the policy (training)        & = & 11.68 \\
Value-function of the non-optimal policy       & = & .    \\
Value-function of the optimal policy (training)& = & 13.63 \\
Rate of optimal policy matches                 & = & 0.31 \\
\hline
\multicolumn{3}{c}{\textbf{New Data Results}} \\
\hline
Value-function of the non-optimal policy (new) & = & .    \\
Value-function of the optimal policy (new)     & = & .    \\
\hline
\end{tabular}
\end{table}

\begin{table}[htbp]
\centering
\caption{Main Results: Risk-averse linear}
\label{tab:main_results_tabL}
\scriptsize
\begin{tabular}{lcl}
\hline
\multicolumn{3}{c}{\textbf{Data Information}} \\
\hline
Number of training observations                    & = & 31,866  \\
Number of used training observations (optimal policy)     & = & 31,866  \\
Number of used training observations (non-optimal policy) & = & .   \\
Number of new observations                         & = & 0  \\
Number of used new observations (optimal policy)   & = & 0  \\
Number of used new observations (non-optimal policy) & = & .   \\
\hline
\multicolumn{3}{c}{\textbf{Policy Information}} \\
\hline
Target variable        & : & net farm output \\
Features               & : & Labor, fixed costs, variable costs, \\
                       &   & subsidy (pillar 2), subsidy (national) \\
Policy variable        & : & treat \\
Number of actions      & = & 3 \\
Actions                & = & \{0, 1, 2\} \\
\hline
\multicolumn{3}{c}{\textbf{Frequencies of the actions (training data)}} \\
\hline
Action 0 & Freq. = 10,876 & Percent = 34.13 \\
Action 1 & Freq. = 10,494 & Percent = 32.93 \\
Action 2 & Freq. = 10,496 & Percent = 32.94 \\
Total    & Freq. = 31,866 & Percent = 100.00 \\
\hline
\multicolumn{3}{c}{\textbf{Training Data Results}} \\
\hline
Value-function of the policy (training)        & = & 7.80 \\
Value-function of the non-optimal policy       & = & .    \\
Value-function of the optimal policy (training)& = & 11.16 \\
Rate of optimal policy matches                 & = & 0.31 \\
\hline
\multicolumn{3}{c}{\textbf{New Data Results}} \\
\hline
Value-function of the non-optimal policy (new) & = & .    \\
Value-function of the optimal policy (new)     & = & .    \\
\hline
\end{tabular}
\end{table}

\begin{table}[htbp]
\centering
\caption{Main Results: Risk-averse quadratic}
\label{tab:main_results_tabQ}
\scriptsize
\begin{tabular}{lcl}
\hline
\multicolumn{3}{c}{\textbf{Data Information}} \\
\hline
Number of training observations                    & = & 31,866  \\
Number of used training observations (optimal policy)     & = & 31,866  \\
Number of used training observations (non-optimal policy) & = & .   \\
Number of new observations                         & = & 0  \\
Number of used new observations (optimal policy)   & = & 0  \\
Number of used new observations (non-optimal policy) & = & .   \\
\hline
\multicolumn{3}{c}{\textbf{Policy Information}} \\
\hline
Target variable        & : & net farm output \\
Features               & : & Labor, fixed costs, variable costs, \\
                       &   & subsidy (pillar 2), subsidy (national) \\
Policy variable        & : & treat \\
Number of actions      & = & 3 \\
Actions                & = & \{0, 1, 2\} \\
\hline
\multicolumn{3}{c}{\textbf{Frequencies of the actions (training data)}} \\
\hline
Action 0 & Freq. = 10,876 & Percent = 34.13 \\
Action 1 & Freq. = 10,494 & Percent = 32.93 \\
Action 2 & Freq. = 10,496 & Percent = 32.94 \\
Total    & Freq. = 31,866 & Percent = 100.00 \\
\hline
\multicolumn{3}{c}{\textbf{Training Data Results}} \\
\hline
Value-function of the policy (training)        & = & 6.34 \\
Value-function of the non-optimal policy       & = & .    \\
Value-function of the optimal policy (training)& = & 11.12 \\
Rate of optimal policy matches                 & = & 0.31 \\
\hline
\multicolumn{3}{c}{\textbf{New Data Results}} \\
\hline
Value-function of the non-optimal policy (new) & = & .    \\
Value-function of the optimal policy (new)     & = & .    \\
\hline
\end{tabular}
\end{table}

\begin{figure}[htbp]
    \centering
    \includegraphics[width=1\textwidth]{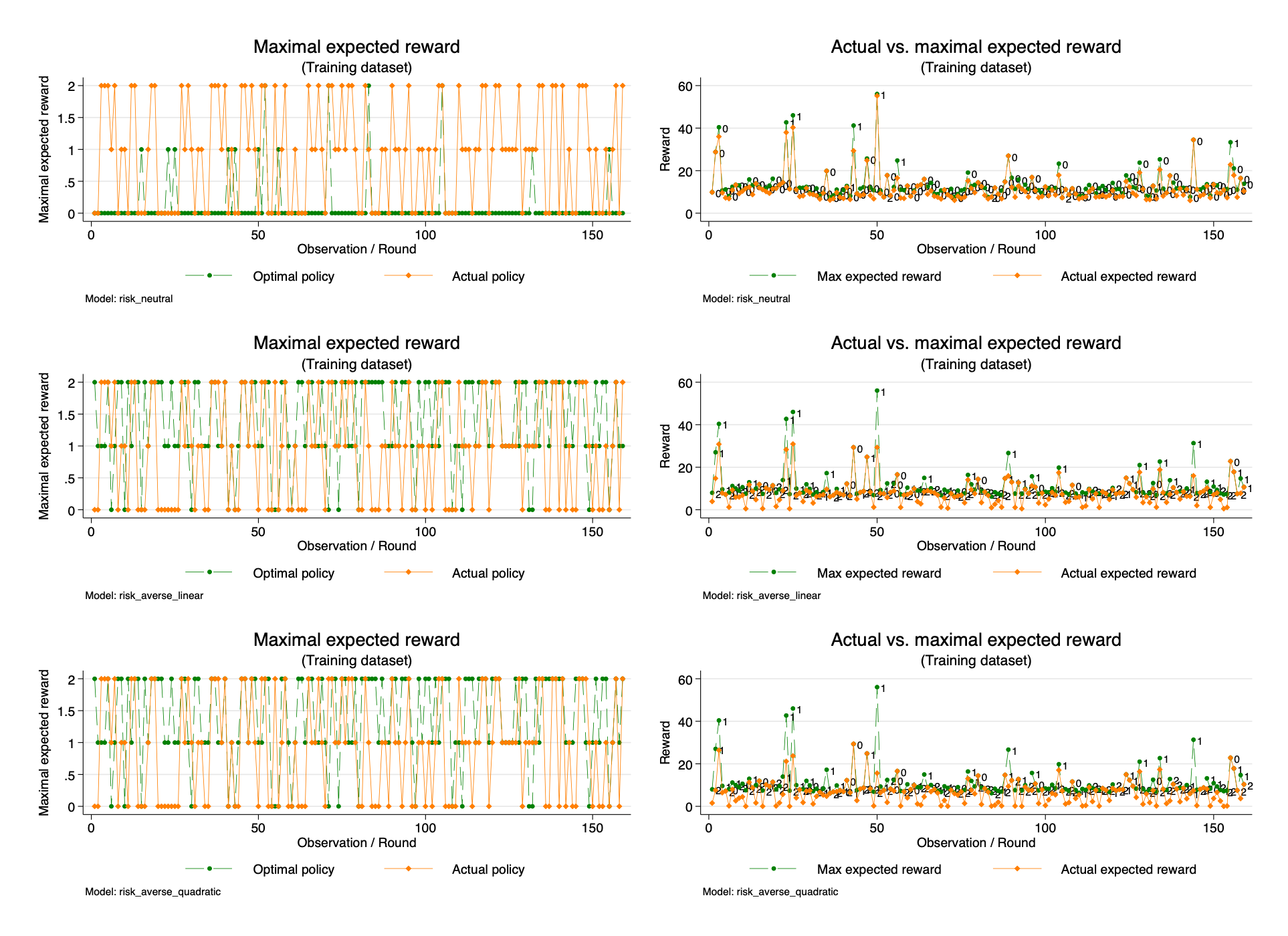}
    \caption{\footnotesize Actual vs. optimal action allocation (left) and actual vs. maximal expected reward (right) according to three different risk preferences. For graphical rendering purposes, each sub-graph has been generated using a random draw of only 0.5\% of the original sample. Plotting the same graphs for the full dataset of 31,866 observations would result in highly cluttered and unreadable images. The sub-sampling affects only the visualization and has no impact on the estimation or interpretation of the results.}
    \label{fig:gr_opl}
\end{figure}

\noindent
Figure~\ref{fig:gr_opl} provides graphical representations of the previous results. It is composed of six sub-figures, arranged in three rows that correspond to the three risk preference specifications considered in the analysis: risk-neutral, risk-averse linear, and risk-averse quadratic. 
Within each row, the left-hand panel contrasts the actual allocation of actions with the counterfactual optimal allocation derived under the corresponding risk preference, while the right-hand panel compares the expected outcome under the actual policy with the maximal expected outcome attainable under the optimal rule. 

For graphical clarity, each sub-figure is based on a random draw of only 0.5\% of the full dataset of 31,866 observations. 
This sub-sampling affects visualization only, ensuring that the figures remain interpretable, while all estimation and inference have been carried out on the complete dataset. 
Plotting the entire sample would result in excessive clutter and substantially reduce readability. 

The figure highlights two consistent patterns. 
First, across all risk preferences, the actual allocation of actions diverges substantially from the counterfactual optimal policy, illustrating systematic inefficiency in the observed assignment mechanism. 
Second, in every case the expected outcome under the optimal rule dominates the realized outcome, with the gap being largest under the risk-neutral criterion and more conservative under risk-averse preferences. 
This provides visual evidence of the welfare gains that can be achieved through optimal policy learning, while also showing how the magnitude of such gains depends on the assumed risk attitude of the policymaker.

\section{Discussion and Conclusions} \label{sec:concl}

This paper develops a risk–adjusted alternative to standard optimal policy learning by importing Roy’s (1952) \emph{safety–first} principle into the treatment assignment problem with observational data. 
On the theory side, we formalize a welfare functional that maximizes the probability that outcomes exceed a socially required threshold and show that the resulting \emph{pointwise} optimal policy ranks treatments by the ratio of conditional means to conditional standard deviations. 
Crucially, the familiar mean-variance rule is not a by-product of quadratic utility or mean–variance preferences; it arises from a probabilistic safety criterion that places explicit weight on downside risk.

On the empirical side, we implement the framework using FADN microdata to evaluate the allocation of CAP subsidies. 
The results consistently indicate that data–driven optimal policies improve upon the realized allocation across all risk attitudes. 
Welfare levels are highest under risk neutrality and become progressively more conservative under risk-averse objectives, reflecting the \textit{social cost} of insuring against uncertainty. 
At the same time, the persistent gap between actual and optimal rules—together with the low rate of policy matches—highlights systematic inefficiencies in current assignment mechanisms.

Consistent with this low match rate, prior work documents that first pillar subsidies have, on average, dampened total factor productivity and delivered mixed (often small) efficiency gains (Rizov et al., 2013), while a substantial share of support is capitalized into land values, shifting benefits toward landowners and weakening equity objectives (Ciaian et al., 2015). Against this backdrop, our risk-adjusted OPL results make the efficiency–insurance trade-off explicit: policies that incorporate downside risk deliver lower mean welfare than risk-neutral benchmarks but greater stability—an effect in line with evidence that direct payments stabilize farm incomes (Ciliberti \& Frascarelli, 2015). Together, these patterns suggest that replacing entitlement-based allocation with transparent, data-driven rules—subject to budget and fairness constraints—could raise cost-effectiveness and targeting precision without abandoning the insurance role of the first pillar.

From a policy perspective, our findings support the integration of risk-sensitive criteria into the design of agricultural support and, more broadly, social programs characterized by outcome volatility. 
Risk-neutral targets maximize average gains, but they can expose beneficiaries to substantial variance; risk-averse targets reduce this exposure at the cost of lower mean performance. 
The proposed safety–first OPL framework makes this efficiency–insurance trade-off explicit and quantifiable, enabling policymakers to tailor decisions to their tolerance for risk and to stakeholder preferences.

The analysis also underscores several limitations that motivate future research. 
First, identification relies on unconfoundedness and overlap, and estimation inherits the usual challenges of finite-sample learning with flexible models; systematic diagnostics and sensitivity analysis (e.g., to residual confounding and model misspecification) are essential. 
Second, while the mean-variance policy arises naturally under location–scale assumptions, alternative risk measures—such as quantile–based welfare, CVaR/expected shortfall, or distributionally robust criteria—deserve dedicated treatment and may be preferable in heavy–tailed or asymmetric settings. 
Third, we focus on static, unconstrained assignment; important extensions include budget/eligibility constraints, fairness or equity requirements, dynamic policies, and multi-objective formulations that balance efficiency with resilience and distributional goals. 

Finally, in this study, we estimated the value function using the Direct Method (also known in program evaluation ad Regression Adjustment). Future steps could enhance robustness by implementing alternative off-policy evaluation techniques, such as Inverse Probability Weighting (IPW) or Doubly Robust (DR) estimators, possibly combined with cross-fitting (Dudík, Langford \& Li, 2011)

In sum, bringing safety–first reasoning to OPL yields a transparent and implementable risk–adjusted rule that is empirically effective and normatively interpretable. 
By clarifying how welfare gains vary with risk attitudes, the framework offers practical guidance for agencies that must navigate uncertainty while pursuing efficient, equitable, and robust policy design.

\section{Acknowledgment}
\label{sec:Acknowledgment} 
This work was supported by: FOSSR (Fostering Open Science in Social Science Research),
funded by the European Union - NextGenerationEU under the NPRR grant agreement MUR IR0000008;
PRIN Project RECIPE (Linking Research Evidence to Policy Impact and Learning: Increasing the Effectiveness of Rural Development Programmes Towards Green Deal Goals), MUR code: 20224ZHNXE.

\end{document}